\documentclass[journal,10pt,onecolumn]{IEEEtran}

\ifCLASSINFOpdf
\usepackage[pdftex]{graphicx}
\else
\fi
\usepackage{cite}
\usepackage{amsmath,amssymb,amsfonts}
\usepackage{algorithmic}
\usepackage{graphicx}
\usepackage{textcomp}

\usepackage{epstopdf}
\usepackage{amsmath,amsthm, amssymb}
\usepackage{threeparttable}
\usepackage{bm}
\usepackage{multirow}
\usepackage{booktabs}
\usepackage{color}
\usepackage{tabularx}
\usepackage{setspace}

\usepackage{verbatim}
\usepackage{stfloats}

\begin{document}
\begin{spacing}{2.0}

\title{{Time-dependent Performance Analysis of the 802.11p-based Platooning Communications Under Disturbance}}

\author{Qiong Wu,~\IEEEmembership{Member,~IEEE}, Hongmei Ge, Pingyi Fan,~\IEEEmembership{Senior Member,~IEEE}, Jiangzhou Wang,~\IEEEmembership{Fellow,~IEEE},
Qiang Fan and Zhengquan Li,~\IEEEmembership{Member,~IEEE}

\thanks{Copyright (c) 2015 IEEE. Personal use of this material is permitted. However, permission to use this material for any other purposes must be obtained from the IEEE by sending a request to pubs-permissions@ieee.org.

This work was supported in part by the National Natural Science Foundation of China under Grant Nos. 61701197 and 62072216, in part by the Beijing Natural Science Foundation under Grant No. 4202030, in part by the 111 Project under Grant No. B12018, in part by the Open Foundation of State key Laboratory of Networking and Switching Technology (Beijing University of Posts and Telecommunications) under Grant No. SKLNST-2020-1-13. (Corresponding author: Qiong Wu)

Qiong Wu and Zhengquan Li are with the School of Internet of Things Engineering, Jiangnan University, Wuxi 214122, China, and also with the State key Laboratory of Networking and Switching Technology, Beijing University of Posts and Telecommunications, Beijing 100876, China (e-mail: \{qiongwu, lzq722\}@jiangnan.edu.cn).

Hongmei Ge is with the  School of Internet of Things Engineering, Jiangnan University, Wuxi 214122, China (hongmeige@stu.jiangnan.edu,cn).

Pingyi Fan is with the Department of Electronic Engineering, Beijing National Research Center for Information Science and Technology, Tsinghua University, Beijing 100084, China (email: fpy@tsinghua.edu.cn).

Jiangzhou Wang is with the School of Engineering and Digital Arts, University of Kent, CT2 7NT Canterbury, U.K. (Email: j.z.wang@kent.ac.uk).

Qiang Fan is with the Department of Electrical and Computer Engineering, New Jersey Institute of Technology, Newark NJ 07102 USA. (Email: qf4@njit.edu).
}

}


\maketitle

\begin{abstract}
Platooning is a critical technology to realize autonomous driving. {Each vehicle in platoons adopts the IEEE 802.11p standard to exchange information through communications to maintain the string stability of platoons. However, one vehicle in platoons inevitably suffers from a disturbance resulting from the leader vehicle acceleration/deceleration, wind gust and uncertainties in a platoon control system, i.e., aerodynamics drag and rolling resistance moment etc. Disturbances acting on one vehicle may inevitably affect the following vehicles and cause that the spacing error is propagated or even amplified downstream along the platoon, i.e., platoon string instability. In this case, the connectivity among vehicles is dynamic, resulting in the performance of 802.11p in terms of packet delay and packet delivery ratio being time-varying. The effect of the string instability would be further deteriorated once the time-varying performance of 802.11p cannot satisfy the basic communication requirement. Unlike the existing works which only analyze the steady performance of 802.11p in vehicular networks, we will focus on the impact of disturbance and construct models to analyze the time-dependent performance of 802.11p-based platooning communications. The effectiveness of the models is validated through simulation results. Moreover, the time-dependent performance of 802.11p is analyzed through numerical results and it is validated that 802.11p is able to satisfy the communication requirement under disturbance.}
\end{abstract}


\IEEEpeerreviewmaketitle

\section{Introduction}
\label{sec1}
In recent years, studying autonomous driving has been paid much attention in order to improve the safety of traveling and energy consumption \cite{Shim2015}.
For the platooning technology, the autonomous vehicles equipped with various sensors are organized into multi-platoons. Each platoon consists of a leader vehicle and several member vehicles. Specifically, the leader vehicle controls the kinematics of the platoon. {Each member vehicle follows the leader vehicle one by one on the same lane.} Actually, the leverage of platooning technology can reduce both the traffic accidents and the cost of fuels \cite{Turri2016}.

{The key task in the vehicle platoon is how to maintain string stability, i.e., keeping the same velocity for each vehicle in a platoon and the reasonable intra-platoon spacing, so that the vehicles in a platoon can move as one entity. The automatic control technology and wireless communication technology are the two main technologies to maintain the string stability of platoon. For the automatic control technology, each vehicle in the platoon is equipped with the autonomous cruise control (ACC) system to sense the spacing to the preceding vehicle \cite{Kayacan2017}. Once the sensed spacing changes, the kinestate of the individual vehicle including the acceleration and velocity is adjusted in real time according to the intelligent driving model (IDM) \cite{Bijan2017}. For the wireless communication technology, vehicles in the platoon communicate with each other to exchange message \cite{Ucar2018}, so that each vehicle can obtain the messages from other vehicles and react accordingly, to maintain the platoon string stability. Here the IEEE 802.11p standard is employed as the communication access protocol in the vehicular environment. In particular, the enhanced distributed channel access (EDCA) mechanism is adopted in order to guarantee different quality of service (QoS) requirement for multi-traffics \cite{Chang2015}.} Specifically, the 802.11p EDCA mechanism employs multiple access categories (ACs) queues with different priorities to transmit different types of safety information. Two typical safety information are event-driven information (e.g., sudden accidents and burst brakes) and periodic information (e.g., the kinestates of vehicles such as moving direction, velocity and acceleration) \cite{Noor2019}. The event-driven information is usually more emergent than the periodic information, and thus is granted higher-priority AC in the 802.11p EDCA mechanism to obtain higher-level QoS.

{However, the autonomous vehicles in platoons inevitably suffer from disturbances resulting from the leader vehicle acceleration/deceleration, wind gust and uncertainties in a platoon control system, i.e., the aerodynamics drag, rolling resistance moment, random non-Gaussian and non-linear noise, and variation of vehicle mass etc., \cite{Guo2011, Liu2014}. Disturbance acting on one vehicle would change its velocity and acceleration which would further change the spacing to the following vehicle, i.e., spacing error. Then each of the following vehicles has to adjust its kinestate according to the IDM model online based on the spacing sensed by the ACC system. The spacing error is propagated or even amplified downstream along multi-platoon, causing platoon string instability. In this case, the connectivity among vehicles is dynamic, resulting in the 802.11p working in time-varying mode. Once the time-varying 802.11p cannot satisfy the basic communication requirement, i.e., vehicles cannot receive the safety information successfully within required time, the resulted string instability would be further deteriorated. Both of them are affected each other. To the time-varying 802.11p, the packet delay (PD) and packet delivery ratio (PDR) are two critical metrics, determining if vehicles can receive the safety information successfully within required time. Therefore, it is critical to model and analyze the time-dependent performance of 802.11p in terms of PD and PDR when a vehicle in a platoon suffers from the disturbance, which is the motivation of our work. Moreover, the dynamic connectivity among vehicles may enhance the non-stationary effects on the performance of 802.11p, which poses a significant challenge to our work.}


{In the literature, many works focused on analyzing the steady performance of 802.11p, i.e., the system is working in the scenarios without time-varying, 
and only a few works analyzed the time-dependent performance of 802.11p. 
To the best of our knowledge, there is no work considering the time-dependent performance of 802.11p in the presence of disturbance in platoons, which motivates us to investigate it.} In this paper, we analyze the {time-dependent} performance of 802.11p when a vehicle in the platoons encounters a disturbance. Our main contributions are summarized as follows.

\begin{itemize}
\item[1)] { We consider the impact of disturbance and construct models to analyze the time-dependent performance of 802.11p-based platooning communications and find that 802.11p is able to satisfy the communication requirement under disturbance.}
\item[2)] Given the initial kinestate of vehicles, we derive the {time-dependent} kinestate of each vehicle after a leader vehicle in the platoons suffers from disturbance based on the IDM model and construct a network connectivity model based on the changing kinestate of each vehicle to reflect the {time-dependent} connectivity among vehicles.
\item[3)] We adopt the pointwise stationary fluid-flow approach (PSFFA) to model the average number of packets of different ACs with the infinite buffer size for a vehicle and further obtain the {time-dependent} average number of packets in different ACs by solving the PSFFA equation.
\item[4)] We adopt a z-domain linear system to describe the access process of the 802.11p EDCA mechanism and derive the mean and variance of the service time based on the connectivity model.
\item[5)] The {time-dependent} performance of 802.11p is studied in terms of the packet delay and packet delivery ratio based on the obtained {time-dependent} average number of packets. Moreover, we validate the effectiveness of models through simulation results and analyze the time-varying performance of 802.11p through numerical results.
\end{itemize}

The rest of the paper is organized as follows. Section \uppercase\expandafter{\romannumeral2} provides a review of related work. Section \uppercase\expandafter{\romannumeral3} describes the system model including the description of scenario and the overview of the 802.11p EDCA mechanism. Section \uppercase\expandafter{\romannumeral4} introduces the models to analyze the {time-dependent} performance of 802.11p for platooning communications under the disturbance. Section \uppercase\expandafter{\romannumeral5} describes the algorithm to calculate the nonstationary performance. Section \uppercase\expandafter{\romannumeral6} validates the models and analyzes the {time-dependent} performance through numerical results. Section \uppercase\expandafter{\romannumeral7} concludes this paper.

\section{Related Work}
\begin{figure*}
\centering
\includegraphics[scale=0.46]{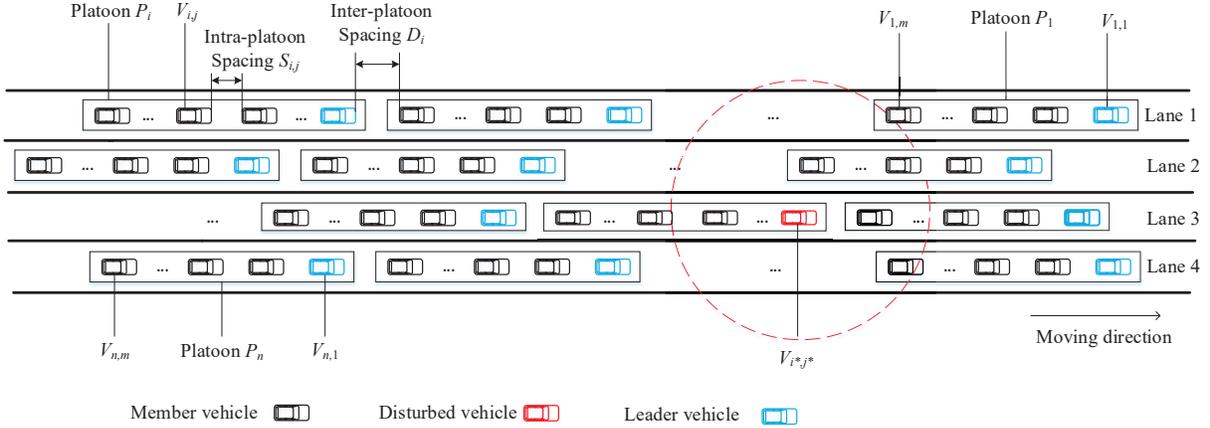}
\caption{Initial platooning scenario.}
\label{fig1}
\end{figure*}

Several wireless access technologies have been used for platooning communication, including IEEE 802.11p, cellular technologies, IEEE 802.11bd, new radio (NR) and infrared communications \cite{Peng2019}. IDifferent from other communication technologies, 802.11p has been widely employed as the communication access protocol in the vehicular environment. In this section, we first review the works about the communication technologies which have been used for platooning communication except the 802.11p, and then the works about the performance analysis of the 802.11p are introduced.

\subsection{Platooning Communication Technologies Except 802.11p}
Some works have focused on the communication technologies for platooning communication except 802.11p.
In \cite{Li2019}, Li \emph{et al.} evaluated the Responsibility-Sensitive Safety (RSS) strategy supported by Long Term Evolution-Vehicle (LTE-V)-based communication. Then, they improved the RSS strategy by considering the uncertainty of transmission time as well as its dependence on vehicle distance and density to prevent collision for automated vehicles. Finally, they analyzed the performance of the proposed strategy under two basic scenarios, i.e., vehicle platoon and approaching vehicle process.
In \cite{Peng2017}, Peng \emph{et al.} proposed a resource allocation approach to provide the timely and successful delivery of LTE-based inter-vehicle information in a multi-platooning scenario, and presented power control scheme to guarantee the transmission rate of each device-to-device link while minimizing the transmission power of each vehicle.
In \cite{Naik2019}, Naik \emph{et al.} studied two radio access technologies (RATs), i.e., IEEE 802.11bd and new radio, which support many advanced vehicular applications such as vehicle platooning, advanced driving, extended sensors and remote driving. They outlined the key objectives of the two RATs, described their salient features and provided an in-depth description of the key mechanisms.
In \cite{Fernandes20120},  Fernandes \emph{et al.} proposed a new inter-vehicle communication technology for platooning systems, i.e., adding infrared (IR) to transmit most of the control data and utilizing dedicated short range communications (DSRC) to fulfill communication purposes.

{\subsection{Performance Analysis of 802.11p}}
Some existing works have studied how to analyze the performance of 802.11p in the steady state.
{In \cite{Yao2019}, Yao \emph{et al.} constructed an analytical model to evaluate the one-hop broadcast performance of IEEE 802.11p, and derive the explicit expressions of the probability distribution, mean and deviation of media access control (MAC) access delay through the proposed analytical model. Besides, they proved that exponential distribution can be a good approximation for the MAC access delay utilizing K-S test.}
In \cite{Noor-A-Rahim2018}, Noor-A-Rahim \emph{et al.} proposed an analytical model to evaluate the performance of 802.11p safety message broadcast for an intersection scenario and then proposed a scheme to improve the performance including the packet reception rates and channel access delay through employing the road side unit (RSU) to rebroadcast the message.
In \cite{Xie2017}, Xie \emph{et al.} proposed a stochastic vehicular traffic model to evaluate the performance of 802.11p-based vehicular ad-hoc networks unicast in terms of the network delay and throughput, and developed a cross-layer optimization method to improve the network performance.
In \cite{Kim2016}, Kim \emph{et al.} proposed an analytical model on the enhanced distributed channel access for the control channel of 802.11p to evaluate the performance in terms of successful delivery probability and delay distribution.
In \cite{Togou2017}, Togou \emph{et al.} proposed a Markovian analytical model that considers the busy channel to evaluate the throughput of the 802.11p EDCA mechanism for vehicle-to-vehicle non-safety applications.
In \cite{Wu2015}, Zheng and Wu proposed two Markov models to analyze the performance of the 802.11p EDCA mechanism. One model was used to describe the contention period after a busy period and the other was used to describe the backoff process for an AC.
In \cite{Qiu2014}, Qiu \emph{et al.} proposed a stochastic traffic model to describe practical vehicular density and developed a model to evaluate the broadcast performance of 802.11p in VANETs. The two models are coupled to predict various network performance metrics including delay, broadcasting efficiency and throughput.
In \cite{Hassan2012}, Hassan \emph{et al.} proposed an analytical model taking into account the hidden terminals and unsaturated situation to predict the performance of 802.11p for vehicle-to-vehicle safety messages with and without retransmission.
In \cite{peng2016}, Peng \emph{et al.} proposed a model to analyze the performance of 802.11p in multi-platoons including the formulations of packet delay, collision probability, transmission attempt probability, packet-dropping probability and network throughput.
In \cite{Han2012}, Han \emph{et al.} proposed an analysis model to analyze the throughput of 802.11p taking into account the contention window and arbitration inter-frame space (AIFS) of different queues, along with internal collision resolution mechanisms under saturated traffic conditions.

 As described above, a lot of work has been conducted to study the performance of 802.11p in steady state, but only a few of works focused on the time-varying performance of 802.11p. In \cite{Xu2015}, Xu \emph{et al.} considered the vehicular movements in a two-way eight-lane freeway scenario and proposed a fluid-flow (FF) differential equation to describe the time-varying performance of transmission queues including PD and PDR when 802.11p standard was adopted in the dynamic VANETs. In \cite{Bilstrup2008}, Bilstrup \emph{et al.} analyzed real-time requirements for vehicular communications in 802.11p through simulations. They demonstrated that the carrier sense multiple access (CSMA) based 802.11p standard was unaccommodated for time-critical traffic safety applications and presented the self-organizing time division multiple access (STDMA) to support real-time application in VANETs. In \cite{Tong2015}, Tong \emph{et al.} adopted stochastic geometry to derive a model which describes real-time the temporal and spatial behavior of 802.11p standard.

{In the literature, many works focused on analyzing the steady performance of 802.11p, i.e., the system is working in the scenarios without time-varying and no work considered the time-dependent performance of 802.11p in the presence of disturbance in platoons, which motivates us to investigate it. The comparison of existing works is given in Table \ref{tab0}.}

\begin{table}
\scriptsize
\caption{Existing Works Comparison}
\label{tab0}
\centering
\begin{tabular}{cccc}
\toprule
\textbf{Research} &\textbf{Steady/Time varying} &\textbf{Platoon} &\textbf{Disturbance}\\
\midrule
Yao \cite{Yao2019} & steady-state & $\times$   & $\times$ \\
\midrule
Noor-A-Rahim \cite{Noor-A-Rahim2018} & steady-state & $\times$   & $\times$ \\
\midrule
Xie \cite{Xie2017} & steady-state & $\times$   & $\times$\\
\midrule
Kim \cite{Kim2016} & steady-state & $\times$   & $\times$\\
\midrule
Togou \cite{Togou2017} & steady-state & $\times$   & $\times$\\
\midrule
Wu \cite{Wu2015} & steady-state & $\times$   & $\times$\\
\midrule
Qiu \cite{Qiu2014} & steady-state & $\times$   & $\times$\\
\midrule
Hassan \cite{Hassan2012} & steady-state & $\times$   & $\times$\\
\midrule
Peng \cite{peng2016} & steady-state & $\surd$   & $\times$\\
\midrule
Han \cite{Han2012} & steady-state & $\times$   & $\times$\\
\midrule
Xu \cite{Xu2015} & time-varying & $\times$   & $\times$\\
\midrule
Bilstrup \cite{Bilstrup2008} & time-varying & $\times$   & $\times$\\
\midrule
Tong \cite{Tong2015} & time-varying & $\times$   & $\times$\\
\midrule
This Paper & time-varying & $\surd$   & $\surd$\\
\bottomrule
\end{tabular}
\end{table}

\section{System Model}
In this section, the platooning scenario is introduced, then the access process of the 802.11p EDCA mechanism is overviewed.
\subsection{Platooning Scenario}
Consider an initial platooning scenario as shown in Fig. \ref{fig1} that there are $n$ platoons moving on a one-way four-lane highway. Initially, the formation of each platoon is stable, i.e., the velocity of each vehicle is ${v_{stb}}$; the intra-platoon spacing and inter-platoon spacing between two adjacent vehicles in platoons are the constant spacing at the equilibrium point of the IDM model \cite{Jia2014}. Vehicles exchange information with each other through the 802.11p EDCA mechanism, which is described in detail in the next subsection. The initial kinestate of each vehicle (including position, velocity and acceleration) are known through communications. The number of vehicles in each platoon is equal to $m$. Let $P_i$ be the $i$-th platoon and $V_{i,j}$ be the $j$-th vehicle in the $i$-th platoon ($1\leq i\leq n$, $1\leq j\leq m$). Each vehicle is equipped with an ACC system at the bumper to sense the distance. In this case, the intra-platoon spacing sensed by $V_{i,j}$, i.e., $S_{i,j}$, is the spacing between the rear of $V_{i,j-1}$ and the bumper of $V_{i,j}$, and the inter-platoon spacing sensed by $V_{i,1}$, i.e., $D_{i}$, is the spacing between the rear of $V_{i-1,m}$ and the bumper of $V_{i,1}$.

\begin{figure}
\centering
\includegraphics[scale=0.38]{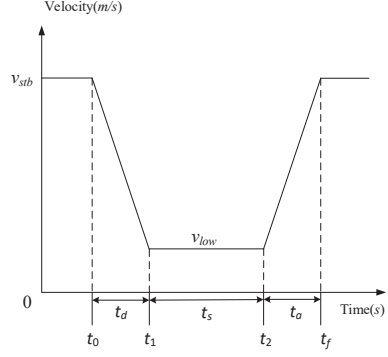}
\caption{Velocity change of a vehicle under disturbance.}
\label{fig2}
\end{figure}

{In the initial scenario as shown in Fig. \ref{fig1}, vehicle $V_{i^*,j^*}$ encounters a disturbance which may be caused by the leader vehicle acceleration/deceleration, wind gust and uncertainties in a platoon control system, i.e., the aerodynamics drag, rolling resistance moment, random non-Gaussian and non-linear noise, and variation of vehicle mass etc., \cite{Guo2011, Liu2014}. To follow the preceding vehicle, the following vehicles need to adjust their velocity according to some models (IDM) once disturbances occur. In this paper, we use the velocity change of a vehicle under disturbance proposed in \cite{Liu2014} and \cite{Fernandes2012} to describe the time-varying kinestate of $V_{i^*,j^*}$, which is shown in  Fig. \ref{fig2}.} Specifically, vehicle $V_{i^*,j^*}$ first decelerates its velocity from $v_{stb}$ to $v_{low}$ for a time duration $t_d$ with a constant negative acceleration, then keeps moving with the constant velocity $v_{low}$ for a time duration $t_s$, finally accelerates from $v_{low}$ to $v_{stb}$ for a time duration $t_a$ with a constant positive acceleration. Each of the following vehicles adopts the IDM model to adjust its velocity and acceleration dynamically based on the sensed distance to follow towards the preceding vehicle.


\subsection{Overview of 802.11p EDCA Mechanism}

In the IEEE 802.11p EDCA mechanism, vehicles adopt multiple transmission queues (i.e., AC queue) with different priorities to access the channel. In this paper, we consider two typical information on the control channel, i.e., event-driven information and periodic information. The event-driven information is transmitted in an AC queue with a high priority, i.e., $AC_0$, while the periodic information is transmitted in an AC queue with a low priority, i.e., $AC_1$. The transmission mode is broadcast. {Similar to most related work \cite{Kim2016},\cite{Yao2013J}, we assume that the physical channel condition is ideal, i.e., the transmission error is mainly caused by data collision and the effect of the physical channel is ignored compared to the data collision.} Packets arrive at $AC_0$ according to the Possion process with arrival rate $\lambda_0(t)$ at time $t$, and arrive at $AC_1$ periodically with arrival rate $\lambda_1(t)$ at time $t$.

When an AC queue in a vehicle has a packet to transmit, it adopts the 802.11p EDCA mechanism to access the channel with its own parameters such as the minimum contention window, maximum contention window and arbitration inter-frame space number. Let $CW_{0,min}$ and $AIFSN_0$ be the minimum contention window and arbitration inter-frame space number for $AC_0$, respectively, and $CW_{1,min}$, $CW_{1,max}$ and $AIFSN_1$ be the minimum contention window, maximum contention window and arbitration inter-frame space number for $AC_1$, respectively. Denote $AIFS_q$ as the time duration of arbitration inter-frame space (AIFS) for $AC_q$ $(q=0,1)$. Here, $AIFS_q$ is calculated as
\begin{equation}
AIFS_q = AIFSN_q \times T_{slot}  + SIFS,
\label{eq1}
\end{equation}
where $T_{slot}$ is the duration of a slot time and $SIFS$ is the duration of short inter-frame spacing.

\begin{figure}
\centering
\includegraphics[scale=0.35]{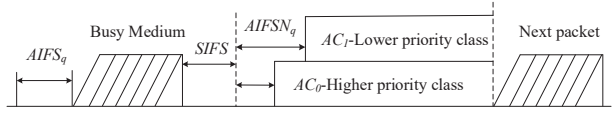}
\caption{802.11p EDCA mechanism.}
\label{fig3}
\end{figure}

The access processes of $AC_0$ and $AC_1$ are described as follows. When $AC_0$ in a vehicle has a packet to transmit, a backoff procedure will be initiated. Specifically, the value of backoff counter is randomly selected from $[0, W_0-1]$, where $W_0$ is the contention window and $W_0=CW_{0,min}+1$, then the value of the backoff counter is decremented by one if the channel is detected to be idle for one time slot. If the channel is busy, the backoff counter will be frozen at this value until the channel keeps idle for $AIFS_0$. When the backoff counter is decremented to $0$, $AC_0$ broadcasts the packet. When $AC_1$ in a vehicle has a packet to transmit, it uses its own parameters to access the channel. The access process of $AC_1$ is the same with that of $AC_0$ if the two AC queues in a same vehicle are not  transmitting simultaneously; otherwise an internal collision occurs at this time and thus $AC_0$ broadcasts its packet while $AC_1$ restarts a backoff procedure with the contention window $W_{1,1}=2(CW_{1,min}+1)$ to retransmit the packet, where $W_{1,r}$ is the contention window of $AC_1$ when the number of retransmissions is $r$ $(r=0,1,\dots)$. For each retransmission, the contention window is doubled accordingly until it reaches $CW_{1,max}+1$, where the number of retransmission is $M$ which is calculated by Eq. \eqref{eq2}. Note that the contention window keeps the value $CW_{1,max}+1$ when the number of retransmissions is larger than $M$ and smaller than the retransmission limit $R$. When the retransmission limit is reached, the packet is dropped, then the value of contention window is reset. The access process of the 802.11p EDCA mechanism is shown in Fig. \ref{fig3}.

\begin{equation}
{M} = {\log _2}\left( {\frac{{CW_{1,\max } + 1}}{{CW_{1,\min}{\rm{ + 1}}}}} \right).\\
\label{eq2}
\end{equation}

\begin{table}
\scriptsize
\caption{The Notations Used in This Paper}
\label{tab1}
\centering
\begin{tabular}{lp{6.6cm}}
\toprule
\textbf{Notation} &\textbf{Definition}\\
\midrule
$a$ &The maximum acceleration.\\
${a_{i,j}}(t)$ &The acceleration of vehicle $V_{i,j}$.\\
$A$ &The $A$ slots which are longer than $AIFS_0$ in $AIFS_1$.\\
$b$ &The comfortable deceleration.\\
${c_q}(t)^2$ &The squared coefficient variation of the service time of $AC_q$ at time $t$.\\
${Ds_q}(t)$ &The variance of service time of $AC_q$.\\
${h_{i,j}^{k,l}}(t)$ &The element of the connectivity model reflects the connectivity between $V_{i,j}$ and $V_{k,l}$ at time $t$.\\
$H(t)$ &The connectivity model at time $t$.\\
$L$ &The length of a vehicle.\\
${L_q}(t)$ &The average number of packets in $AC_q$ at time $t$.\\
$\mathop {L_q}\limits^. (t)$ & The change rate of average number of packets in $AC_q$ at time $t$.\\
$m$ &The number of vehicles in each platoon.\\
$M$ &The retransmission limit of $AC_1$.\\
$n$ &The total number of platoons.\\
$N$ &The total number of vehicles in the scenario.\\
${N_{tr}}(t)$ &The number of vehicles within transmission range of the target vehicle at time $t$.\\
${p_{s}^{q}}(t)$ &The probability that the channel is sensed busy for $AC_q$ at time $t$.\\
${p_{a}^{q}}(t)$ &The packet arrival probability of $AC_q$ at time $t$.\\
${P_{ij,kl}^{c}}(t)$ &The collision probability when vehicle $V_{i,j}$ transmits to vehicle $V_{k,l}$ at time $t$.\\
${P_{ij,kl}^{exposed}}(t)$ &The collision probability caused by the exposed terminal at time $t$.\\
${P_{ij,kl}^{hidden}}(t)$ &The collision probability caused by the hidden terminal at time $t$.\\
${PD_q}(t)$ &The packet delay of $AC_q$ at time $t$.\\
${PDR_q}(t)$ &The packet delivery ratio of $AC_q$ at time $t$.\\
$R$ &The retransmission limit.\\
$R_d$ &The date rate.\\
$R_b$ &The basic rate.\\
$R_{tr}$ &The transmission range of each vehicle.\\
$s_{0}$ &The minimum intra-platoon spacing.\\
${S_{i,j}}(t)$ &The intra-platoon spacing between $V_{i,j}$ and $V_{i,j-1}$ at time $t$.\\
${S_{i,j}^{*}}(t)$ &The desired intra-platoon spacing between $V_{i,j}$ and $V_{i,j-1}$.\\
$T_{h}$ &The desired time headway.\\
$T_{tr}$ &The transmission time.\\
$T_{slot}$ &The average length of a slot time.\\
${Ts_q}(t)$ &The mean of service time of $AC_q$.\\
$v_{0}$ &The maximum speed.\\
${v_{i,j}}(t)$ &The velocity of $V_{i,j}$ at time $t$.\\
$V_{i,j}$ &The $jth$ vehicle in $ith$ platoon.\\
$W_0$ &The initial contention window size of $AC_0$.\\
$W_{1,r}$ &The contention window of $AC_1$ when the number of retransmission is $r$.\\
${x_{i,j}}(t)$ &The abscissa of the position of $V_{i,j}$ at time $t$.\\
${y_{i,j}}(t)$ &The ordinate of the position of $V_{i,j}$ at time $t$.\\
$\Delta{v_{i,j}}(t)$ &The velocity difference between $V_{i,j-1}$ and $V_{i,j}$ at time $t$.\\
$\Delta t$ &The update time interval.\\
${\lambda_q}(t)$ &The packet arrival rate at time $t$.\\
${\mu_q}(t)$ &The average service rate of $AC_q$ at time $t$.\\
${\rho_q}(t)$ &The server utilization of $AC_q$ at time $t$.\\
${\sigma_q}(t)$ &The transmission probability of $AC_q$ at time $t$.\\
$\delta $ &The propagation delay.\\
\bottomrule
\end{tabular}
\end{table}
\section{Analytical Models}
In this section, we construct models to analyze the time-varying performance of 802.11p for a target vehicle $V_{i,j}$ once a leader vehicle $V_{i^*,j^*}$ in the platoons suffers from the disturbance. {Specifically, we first construct a connectivity model to reflect the time-varying connectivity among vehicles and derive the time-dependent kinestate of each vehicle to determine the network connectivity model.}
Then we use the pointwise stationary fluid-flow approximation method to model the dynamic queuing behavior of different ACs and derive the time-varying average number of packets in each AC. {Afterwards, the time-varying average number of packets in each AC is determined by the corresponding mean and variance of the service time of each AC.}
Finally, we derive the time-varying performance of each AC for the target vehicle in terms of the packet delay and packet delivery ratio according to the average number of packets. The notations used in this section are summarized in Table \ref{tab1}.
\subsection{Connectivity Model}
In this subsection, a time-varying connectivity model is constructed to reflect the dynamic connectivity among vehicles after each small time interval $\Delta t$. Since the number of platoons is $n$ and the number of vehicles in each platoon is $m$, the number of vehicles in the scenario $N$ becomes $m \times n$. Thus the connectivity model at time $t$ is expressed as an $N \times N$ square matrix $H(t)$, i.e.,
\begin{equation}
H(t) = \left[ {\begin{array}{*{20}{c}}
   {{h_{1,1}^{1,1}}\left( t\right)}  & \cdots & {{h_{1,1}^{k,l}}\left( t\right)} & \cdots & {{h_{1,1}^{n,m}}\left( t\right)} \\
   \vdots & \vdots & \vdots &\vdots \\
   {{h_{i,j}^{1,1}}\left( t\right)} & \cdots & {{h_{i,j}^{k,l}}\left( t\right)} & \cdots & {{h_{i,j}^{n,m}}\left( t\right)}\\
   \vdots & \vdots & \vdots &\vdots \\
   {{h_{n,m}^{1,1}}\left( t\right)} & \cdots & {{h_{n,m}^{k,l}}\left( t\right)} & \cdots & {{h_{n,m}^{n,m}}\left( t\right)} \\
\end{array}}\right],
\label{eq3}
\end{equation}
where ${h_{i,j}^{k,l}(t)}$ denotes whether the $l$-th vehicle in the $k$-th platoon $V_{k,l}$ can connect with the $j$-th vehicle in the $i$-th platoon $V_{i,j}$. Specifically, ${h_{i,j}^{k,l}(t)}=1$ means that $V_{k,l}$ is within the transmission range, i.e., the effective communication distance with high transmission quality, i.e., low bit error. ${h_{i,j}^{k,l}(t)}=0$ means that $V_{k,l}$ is not within the transmission range of  $V_{i,j}$. We assume that  the transmission range  {(the radius is denoted as $R_{tr}$)} of each vehicle is the same, the matrix $H(t)$ is symmetric. The value of ${h_{i,j}^{k,l}(t)}$ can be calculated based on positions of $V_{i,j}$ and $V_{k,l}$. For ease of representing the position of each vehicle, we introduce a rectangular coordinate system, where the center of the disturbed vehicle is the origin and the moving direction of vehicles is the direction of the X-Axis. In this case, positions of $V_{i,j}$ and $V_{k,l}$ at time $t$ can be expressed as (${x_{i,j}}(t)$, ${y_{i,j}}(t)$) and (${x_{k,l}}(t)$, ${y_{k,l}}(t)$), respectively, and thus ${h_{i,j}^{k,l}(t)}$ can be calculated as
\begin{equation}
h_{i,j}^{k,l}(t)\! =\!\! \left\{\! \begin{array}{l}
 \!1,if \sqrt {{{\left[ {{x_{i,j}}(t)\! -\! {x_{k,l}}(t)} \right]}^2} \!+\! {{\left[ {{y_{i,j}}(t) \!- \!{y_{k,l}}(t)} \right]}^2}} \! \le\! {R_{tr}} \\
\!0,others. \\
 \end{array} \right.
\label{eq4}
\end{equation}


Since vehicles are moving along the direction of the X-Axis, ${y_{i,j}}(t)$ and ${y_{k,l}}(t)$ are not time-varying. Therefore, ${x_{i,j}}(t)$ and ${x_{k,l}}(t)$ need to be further derived to determine ${h_{i,j}^{k,l}(t)}$.

For simplicity, assuming that each vehicle follows the uniformly accelerated motion within the time interval $\Delta t$, ${x_{i,j}}(t)$ can be calculated through iterations given the initial kinestate of each vehicle. Denote $v_{i,j}(t)$ and $a_{i,j}(t)$ as the velocity and acceleration of $V_{i,j}$ at time $t$, respectively. According to the uniformly accelerated motion, ${x_{i,j}}(t)$ can be expressed as
\begin{equation}
{x_{i,j}}(t)\! =\! {x_{i,j}}(t \!-\! \Delta t) \!+\! {v_{i,j}}(t\! -\! \Delta t)\Delta t + \frac{1}{2}{a_{i,j}}(t \!-\! \Delta t)\Delta {t^2} + o(\Delta {t^2}).
\label{eq5}
\end{equation}

In Eq. \eqref{eq5},  we need to derive $x_{i,j}(t - \Delta t)$, $v_{i,j}(t - \Delta t)$ and  $a_{i,j}(t - \Delta t)$. {$x_{i,j}(t - \Delta t)$ can be derived by iterations with the kinestate at time $t - 2\Delta t$, i.e., $x_{i,j}(t - 2\Delta t)$, $v_{i,j}(t - 2\Delta t)$ and  $a_{i,j}(t - 2\Delta t)$, according to the uniformly accelerated motion.}



Meanwhile, according to the uniformly accelerated motion equation, $v_{i,j}(t-\Delta t)$ can be expressed as
\begin{equation}
v_{i,j}(t- \Delta t) = v_{i,j}(t-2\Delta t) + a_{i,j}(t-2\Delta t)\Delta t+ o(\Delta {t^2}).\\
\label{eq7}
\end{equation}

{Once disturbance occurs, it will cause the acceleration of $V_{i^*,j^*}$ while the following vehicles also need to adjust their accelerations according to the IDM model. Thus the accelerations of $V_{i^*,j^*}$ and other vehicles at time $t-\Delta t$ can be derived, respectively, as follows.}

\subsubsection{Disturbed Vehicle}
{The velocity change of disturbed vehicle is shown in Fig. \ref{fig2}.} Let $t_0$ be the initial time that  $V_{i,j}$ suffers from the disturbance, $t_1$ be the time that $V_{i,j}$ begins to move with  velocity $v_{low}$, $t_2$ be the time that $V_{i,j}$ begins to accelerate and $t_f$ be the time that $V_{i,j}$ finishes the acceleration. Note that $t_1=t_0+t_d$, $t_2=t_0+t_d+t_s$ and $t_f=t_0+t_d+t_s+t_a$. Thus the acceleration of $V_{i,j}$ is given by
\begin{equation}
a_{i, j}(t-\Delta t)=\left\{\begin{array}{ll}{\frac{v_{\text {low}}-v_{stb}}{t_{d}},} & {t-\Delta t \in\left[t_{0}, t_{1}\right]} \\ {0,} & {t-\Delta t \in\left[t_{1}, t_{2}\right]} \\ {\frac{v_{stb}-v_{low}}{t_{a}},} & {t-\Delta t \in\left[t_{2}, t_{f}\right]}\end{array}\right.
\label{eq8}
\end{equation}

\subsubsection{Following Vehicles}
The accelerations of the following vehicles change in real time according to the IDM model,  thus the acceleration of vehicle $V_{i.j}$ at time $t-\Delta t$ is given by \cite{Treiber2000}
\begin{equation}
{a_{i,j}}(t\! -\! \Delta t) \!=\! a\left[ {1 \!-\! {{\left( {\frac{{{v_{i,j}}(t\! -\! \Delta t)}}{{{v_0}}}} \right)}^4} \!-\! {{\left( {\frac{{S_{i,j}^*(t \!-\! \Delta t)}}{{{S_{i,j}}(t \!-\! \Delta t)}}} \right)}^2}} \right],
\label{eq9}
\end{equation}
where $a$ is the maximum acceleration; $v_0$ is the maximum velocity; $S_{i,j}(t-\Delta t)$ is the intra-platoon spacing between $V_{i,j}$ and $V_{i,j-1}$ at time $t-\Delta t$; $S_{i,j}^*(t-\Delta t)$ is the desired intra-platoon spacing between $V_{i,j}$ and $V_{i,j-1}$. In Eq. \eqref{eq9}, $a$ and $v_0$ are given, and we can get ${v_{i,j}(t-\Delta t)}$ by Eq. \eqref{eq7}. According to the definition of $S_{i,j}(t-\Delta t)$, it can be calculated as
\begin{equation}
\begin{split}
S_{i,j}(t\!-\!\Delta t) \!=\! x_{i,j-1}(t\!-\!\Delta t)\! -\! x_{i,j}(t\!-\!\Delta t)\! -\! L\!+\! o(\Delta {t^2}),\\
\end{split}
\label{eq10}
\end{equation}
where $L$ is the length of vehicle; $x_{i,j-1}(t-\Delta t)$ and $x_{i,j}(t-\Delta t)$ are calculated according to Eq. \eqref{eq5}. Meanwhile, $S_{i,j}^*(t-\Delta t$) is calculated according to the IDM model, i.e.,
\begin{equation}
\begin{split}
S_{i,j}^*(t - \Delta t)& = {s_0} + {v_{i,j}}(t - \Delta t){T_h} \\
&+ \frac{{{v_{i,j}}(t - \Delta t)\Delta {v_{i,j}}(t - \Delta t)}}{{2\sqrt {ab} }} + o(\Delta {t^2}).
\end{split}
\label{eq11}
\end{equation}
where $s_0$ is the minimum intra-platoon spacing and $T_h$ is the desired headway, the value of which is different for leader vehicle and member vehicles. $\Delta v_{i,j}(t-\Delta t)$ is the difference of the velocity between $V_{i,j}$ and $V_{i,j-1}$ at time $t-\Delta t$. According to the definition of $\Delta v_{i,j}(t-\Delta t)$, it is calculated as
\begin{equation}
\begin{split}
\Delta v_{i,j}(t-\Delta t){\rm{ }} &= v_{i,j}(t-\Delta t) - v_{i,j-1}(t-\Delta t)+o(\Delta {t^2}),
\end{split}
\label{eq12}
\end{equation}
where $v_{i,j}(t-\Delta t)$ and $v_{i,j-1}(t-\Delta t)$ are calculated according to Eq. \eqref{eq7}.

From the description above, {$a_{i,j}(t - \Delta t)$ for the disturbed vehicle is obtained according to Eq. \eqref{eq8}.} On the other hand, if vehicle $V_{i,j}$ is the following vehicle, we can derive $a_{i,j}(t - \Delta t)$ according to Eqs. \eqref{eq9}-\eqref{eq12}. Therefore, $x_{i,j}(t)$ can be calculated according to Eq. \eqref{eq5} based on the obtained $x_{i,j}(t-\Delta t)$, $v_{i,j}(t-\Delta t)$ and $a_{i,j}(t - \Delta t)$. As a result, we can derive ${h_{i,j}^{k,l}(t)}$ according to Eq. \eqref{eq4}. Each element of  $H(t)$ is calculated through the method mentioned above, thus the matrix $H(t)$ reflecting connectivity among vehicles at time $t$ is obtained.

Owing to the time-varying connectivity model, the average number of packets changes non-stationarily and thus affects the {time-dependent} performance of 802.11p.

\subsection{Dynamic Behavior of Transmission Queue}

In this subsection, the pointwise stationary fluid-flow approximation method is adopted to model the non-stationary average number of packets of different transmission queues at time $t$ for the target vehicle. Specifically, the FF model is adopted to represent the relationship between the change rate of the average number of packets and the service utilization at time $t$ in each transmission queue by a differential equation. Then, we can further study the relationship between the average number of packets and the service utilization based on the pointwise stationary approximation (PSA) method. Thus we can obtain the approximate differential equation of the average number of packets, which is referred to as PSFFA equation. As a result, the average number of packets of each transmission queue at time $t$ can be calculated by solving the PSFFA equation numerically.






Each transmission queue of the target vehicle is considered as a first-come first-service (FCFS) single-server infinite-capacity queuing system. According to the FF model, the difference between the average arrival rate and departure rate of packets is the change rate of average number of packets \cite{Xu2015}, thus for $AC_q$ we have
\begin{equation}
\mathop {L_q}\limits^. (t) =  - \mu _q(t)\rho _q(t) + \lambda _q(t),
\label{eq13}
\end{equation}
where ${L_q}(t)$ is the average number of packets in $AC_q$ at time $t$; $\mathop {L_q}\limits^. (t)$ is the change rate of ${L_q}(t)$; $\lambda _q(t)$ and $\mu _q(t)$ are the average arrival rate and service rate of packets in $AC_q$, respectively, $\rho_q(t)$ is the server utilization of $AC_q$ at time $t$, thus the packet departure rate is $\mu _q(t)\rho _q(t)$. Note that in Eq. \eqref{eq13}, while $\lambda _q(t)$ is given at time $t$, $\mu _q(t)$ and $\rho _q(t)$ need to be further determined.


Next we will employ the PSA method to derive $\rho _q(t)$. We assume that the service process, i.e., access process, of $AC_q$ follows a general distribution and its service time is independent and equivalently distributed at time $t$ with mean $Ts_q(t)$ and variance $D{s_q}(t)$, where $Ts_q(t)=\frac{1}{{{\mu _q}}(t)}$. Moreover, packets arrive at $AC_0$ according to a Poisson process with the arrival rate $\lambda _0(t)$, while packets arrive at $AC_1$ periodically with the arrival rate $\lambda _1(t)$. Therefore, the transmission queue of $AC_0$ is modeled as an $M/G/1$ queue, while the transmission queue of $AC_1$ follows a $D/G/1$ queue according to the queuing theory \cite{Medhi2002}.

For an $M/G/1$ queue, the relationship between the average number of packets and server utilization can be derived according to the Pollaczek-Khintchine (P-K) formula \cite{Van2006}
 \begin{equation}
{L_0}(t) = {\rho _0}(t) + \frac{{{\rho _0}{{(t)}^2}(1 + {c_0}{{(t)}^2})}}{{2(1 - {\rho _0}(t))}},
\label{eq14}
\end{equation}
where ${c_0(t)^2}$ is the squared coefficient variation of the service time of $AC_0$. Based on Eq. \eqref{eq14}, $\rho _0(t)$ can be derived as
\begin{equation}
{\rho _0}(t) = \frac{{{L_0}(t) + 1 - \sqrt {{L_0}{{(t)}^2} + 2{c_0}{{(t)}^2}{L_0}(t) + 1} }}{{1 - {c_0}{{(t)}^2}}}.
\label{eq15}
\end{equation}

For a $D/G/1$ queue, it is difficulty to obtain the closed-form solution of the average number of packets. According to \cite{KLB1976}, we employ the Kramer and Lagbenbach-Belz (KLB) formula to approximately as
\begin{equation}
{L_1}(t) \approx {\rho _1}(t) + \frac{{{\rho _1}{{(t)}^2}{c_1}{{(t)}^2}{e^{\frac{{ - 2(1 - {\rho _1}(t))}}{{3{\rho _1}(t){c_1}{{(t)}^2}}}}}}}{{2(1 - {\rho _1}(t))}},
\label{eq16}
\end{equation}
where ${c_1(t)^2}$ is the squared coefficient variation of $AC_1$ at time $t$. Since it is difficult to derive the closed-form $\rho _1(t)$ based on Eq. \eqref{eq16}, the curve fitting approach is adopted to determine the polynomial relationship between $\rho _1(t)$ and $L_1(t)$. Specifically, we numerically determine a set of data pair with the element $(L_1(t),\rho _1(t))$ under a given ${c_1(t)}$ based on which the curve fitting method is employed to find the coefficients of the polynomial related to the given parameter ${c_1(t)}$, i.e., $a_0(c_1(t))$, $a_1(c_1(t))$, \ldots, $a_n(c_1(t))$ in the following equation,
\begin{equation}
\begin{split}
{\rho _{\rm{1}}}(t) &= {a_n}({c_1}(t)){L_1}{(t)^n} + {a_{n - 1}}({c_1}(t)){L_1}{(t)^{n - 1}}\\
&+ \ldots + {a_1}({c_1}(t)){L_1}(t) + {a_0}({c_1}(t)).
\end{split}
\label{eq17}
\end{equation}

Substituting Eqs. \eqref{eq15} and \eqref{eq17} into Eq. \eqref{eq13}, we can obtain the PSFFA equation for $AC_0$ and $AC_1$, respectively, i.e.,
\begin{equation}
\begin{split}
\mathop {L_0}\limits^. (t) &= - \frac{{L_0(t) + 1 - \sqrt {L_0{{(t)}^2} + 2c_0{{(t)}^2}L_0(t) + 1} }}{{1 - c_0{{(t)}^2}}}\\
& \times \mu _0(t) + \lambda _0(t).
\end{split}
\label{eq18}
\end{equation}

\begin{equation}
\begin{split}
\mathop {{L_1}}\limits^. (t) &=  - [{a_n}({c_1}(t)){L_1}{(t)^n} + {a_{n - 1}}({c_1}(t)){L_1}{(t)^{n - 1}} + \ldots \\
&+ {a_1}({c_1}(t)){L_1}(t) + {a_0}({c_1}(t))] \times {\mu _1}(t) + {\lambda _1}(t).
\end{split}
\label{eq19}
\end{equation}


In Eqs. \eqref{eq18} and \eqref{eq19}, the arrival rate $\lambda _q(t)$ $(q=0,1)$ is a given constant at time $t$, thus $L_0(t)$ and $L_1(t)$ are determined by both $c_q(t)$ and $\mu _q(t)$. According to the definition of the squared coefficient variation of service time, ${c_q(t)^2}$ is calculated as
 \begin{equation}
 {c_q(t)^2} = \frac{Ds_q(t)}{Ts_q(t)^2},
 \label{eq20}
 \end{equation}
From Eq. \eqref{eq20}, we can see that the $c_q(t)$ is dependent on the mean of the service time $Ts_q(t)$ and the variance of the service time $Ds_q(t)$. Moreover, the relationship between $\mu _q(t)$ and $Ts_q(t)$ is expressed as

 \begin{equation}
 \mu _q(t)=\frac{1}{Ts_q(t)}.
 \label{eq20'}
 \end{equation}

 Therefore, in order to determine $c_q(t)$ and $\mu _q(t)$, we need to further determine $Ts_q(t)$ and $Ds_q(t)$.

\subsection{Mean and Variance of Service Time}
In this subsection, the mean and variance of service time, i.e., $Ts_q(t)$
 and $Ds_q(t)$, are further derived. The service time is the access delay incurred by the access process of the 802.11p EDCA mechanism. Since the access process consists of both the
 transmission procedure and backoff procedure, the access delay is the summation of a
 transmission time and the time duration of each slot in the backoff procedure, where each slot may be occupied by a transmission or an idle slot. We adopt the z-domain linear system in \cite{Yao2013} which is composed of the  probability generating function (PGF) of the transmission time (${TR(z)}$) and the PGF of the average time of each slot (${H(z)}$) to describe the access procedure of the 802.11p EDCA mechanism, then the mean and variance of the service time are calculated through the PGF approach.

 According to the z-domain linear system, we have
 \begin{equation}
\left\{ \begin{array}{l}
P_{Ts}^0(z)\! =\! \frac{{TR(z)}}{{{W_0}}}\!\sum\limits_{h = 0}^{{W_0} - 1} {{{\left[ {{H_0}(z)} \right]}^h}}\\
 P_{Ts}^1(z)\! =\! (1 \!-\! \sigma _0(t))TR(z)\!\sum\limits_{h = 0}^{{R}} {\!\left[ {{{(\sigma _0(t))}^h}\!\prod\limits_{r = 0}^h {G_{1,r}(z)} }\! \right]}  \\
  + {(\sigma _0(t))^{{R} + 1}}\!\prod\limits_{r = 0}^{{R}} {G_{1,r}(z)}  \\
 \end{array} \right.,
\label{eq21}
\end{equation}
 where $P_{Ts}^q(z)$ is the PGF of the service time of $AC_q$; $G_{1,r}(z)$ is the PGF of the backoff time of $AC_1$ when the number of retransmissions is $r$; $H_{q}(z)$ is the PGF of the average time of each slot in $AC_q$ and ${\sigma _q}(t)$ is the transmission probability of $AC_q$ at time $t$.

Moreover, the relationship between $G_{1,r}(z)$ and  $H_{q}(z)$ can be obtained according to the system and is expressed as
\begin{equation}
{G_{1,r}}(z) = \left\{ {\begin{array}{*{20}{c}}
   {\frac{1}{{{W_{1,r}}}}\sum\limits_{h = 0}^{{W_{1,r}} - 1} {{{\left[ {{H_1}(z)} \right]}^h},r \in [0,M{\rm{ - 1}}]} }  \\
   {\frac{1}{{{W_{1,M}}}}\sum\limits_{h = 0}^{{W_{1,M}} - 1} {{{\left[ {{H_1}(z)} \right]}^h},r \in [M,R]} }  \\
\end{array}} \right..
\label{eq22}
\end{equation}

Substituting Eq. \eqref{eq22} into Eq. \eqref{eq21}, ${P_{Ts}^{q}}(z)$ becomes a function of $TR(z)$, $H_{q}(z)$, ${\sigma _0}(t)$. Meanwhile, according to the PGF approach, the mean of the service time $Ts_q(t)$ is calculated as
\begin{equation}
T{s_q}(t) = {\left. {\frac{{dP_{Ts}^{q}(z)}}{{dz}}} \right|_{z = 1}},
\label{eq23}
\end{equation}
and the variance of service time ${Ds_{q}}(t)$ is calculated as

\begin{equation}
D{s_q}(t) = {\left. {\left[ {\frac{{{d^2}P_{Ts}^q(z)}}{{d{z^2}}} + \frac{{dP_{Ts}^q(z)}}{{dz}} - {{(\frac{{dP_{Ts}^q(z)}}{{dz}})}^2}} \right]} \right|_{z = 1}}.
\label{eq24}
\end{equation}

In other words, $Ts_q(t)$ and ${Ds_{q}}(t)$ are dependent on $TR(z)$, $H_{q}(z)$ and ${\sigma _0}(t)$.

Assuming packet size is the same for all ACs, the transmission time can be calculated as
\begin{equation}
{T_{tr}} = \frac{{PH{Y_H}}}{{{R_b}}} + \frac{{MA{C_H} + E[P]}}{{{R_d}}} + \delta,
\label{eq25}
\end{equation}
where $PHY_{H}$ and $MAC_{H}$ are the header length in the physical and MAC layer, respectively; $E[P]$ is the packet size; $R_b$ is the basic rate; $R_ d$ is the data rate; $\delta$ is the propagation delay. As all the parameters in Eq. \eqref{eq25} are predefined, the transmission time ${T_{tr}}$ is deterministic and thus $TR(z)$ is calculated as
\begin{equation}
{TR(z)} = z^{T_{tr}}.
\label{eq26}
\end{equation}

According to the backoff procedure, if the channel is sensed idle for a slot, the time duration of the idle slot is $T_{slot}$. Once the channel is sensed to be busy, the slot would be frozen for a duration $T_{tr}$ until the channel keeps being idle for $AIFS_q$. Let ${p_{s}^q}(t)$ be the probability that the channel is sensed busy at $AC_q$, i.e., the backoff frozen probability at $AC_q$. Therefore, the PGF of average time of each slot for $AC_q$ is calculated as
\begin{equation}
{H_q}(z) = (1 - p_s^q(t)){z^{{T_{slot}}}} + p_s^q(t){z^{{T_{tr}} + AIF{S_q}}},q = 0,1,
\label{eq27}
\end{equation}
where $AIFS_1$ is longer than $AIFS_0$ and the relationship between them is expressed as
\begin{equation}
AIF{S_1} = AIF{S_0} + A \times {T_{slot}}.
\label{eq28}
\end{equation}
For $AC_0$, ${p_{s}^{0}}(t)$ is the probability that the channel is sensed busy in a slot after it keeps idle for $AIFS_0$. For $AC_1$, while as shown in Fig. \ref{fig3}, $AC_1$ needs to detect channel for $A$ more slots than $AC_0$, thus ${p_s^{1}}(t)$ is the probability that the channel is sensed busy at least in one slot within $A+1$ slots. Denote $N_{tr}(t)$ as the number of vehicles at time $t$ within the transmission range of the target vehicle. Thus, ${p_{s}^{q}}(t)$ is calculated as
\begin{equation}
\left\{ \begin{array}{l}
 {p_{s}^0}(t)\!=\! 1 \!-\! {(1 - {\sigma _0}(t))^{{N_{tr}}(t) - 1}}{(1 - {\sigma _1}(t))^{{N_{tr}}(t)}} \\
 {p_{s}^1}(t)\!=\! 1 \!-\! {[{(1 - {\sigma _0}(t))^{{N_{tr}}(t)}}{(1 - {\sigma _1}(t))^{{N_{tr}}(t) - 1}}]^{{A} + 1}} \\
 \end{array} \right.,
\label{eq29}
\end{equation}
where $N_{tr}(t)$ is determined by the network connectivity model $H(t)$ and thus can be calculated as
\begin{equation}
N_{tr}(t) = \sum\limits_{{k} = 1}^n {\sum\limits_{l = 1}^m {{h_{i,j}^{{k,l}}(t) }}}.
\label{eq30}
\end{equation}

In Eq. \eqref{eq29}, ${\sigma _q}(t)$ is calculated according to Eq. \eqref{eq31} \cite{Yao2013}, which is shown at the bottom of this page. In Eq. \eqref{eq31}, ${\rho _q}(t)$ is the server utilization of $AC_q$ at time $t$, ${p_{a}^{q}}(t)$ is the packet arrival probability of $AC_q$ at time $t$. Since the packet arrival at $AC_0$ follows a Poission process with arrival rate ${\lambda_0}(t)$ and packets arrive at $AC_1$ periodically with arrival rate ${\lambda_1}(t)$ at time $t$, the packet arrival probability is calculated as
\newcounter{TempEqCnt}
\setcounter{TempEqCnt}{\value{equation}}
\setcounter{equation}{30}

\begin{equation}
\left\{ \begin{array}{l}
 {\sigma _{\rm{0}}}(t) = {\left[ {\frac{{{W_0} + 1}}{{2\left( {1 - p_s^0(t)} \right)}} + \frac{{1 - {\rho _0}(t)}}{{p_a^0}}} \right]^{ - 1}} \\
 {\sigma _1}(t) = \frac{{1 - {\sigma _0}{{(t)}^{R + 1}}}}{{1 - {\sigma _0}(t)}}\left[ {\frac{{1 - {\sigma _0}{{(t)}^{R + 1}}}}{{1 - {\sigma _0}(t)}} + \frac{{{W_{1,0}} - 1}}{{2\left( {1 - p_s^1(t)} \right)}} + \frac{{{W_{1,0}}{\sigma _0}(t)\left[ {1 - {{\left( {2{\sigma _0}(t)} \right)}^M}} \right]}}{{\left( {1 - p_s^1(t)} \right)\left( {1 - 2{\sigma _0}(t)} \right)}}} \right.{\left. { + \frac{{{2^{M - 1}}{W_{1,0}}{\sigma _0}{{(t)}^{M + 1}}(1 - {\sigma _0}{{(t)}^{R - M}})}}{{\left( {1 - p_s^1(t)} \right)\left( {1 - {\sigma _0}(t)} \right)}} + \frac{{1 - {\rho _1}(t)}}{{p_a^1}}} \right]^{ - 1}} \\
 \end{array} \right..
\label{eq31}
\end{equation}

\setcounter{equation}{\value{TempEqCnt}}

\setcounter{equation}{31}
\begin{equation}
\left\{ \begin{array}{l}
 p_a^0(t) = \sum\limits_{k = 1}^\infty  {\frac{{{{({\lambda _0}(t){T_{slot}})}^k}}}{{k!}}{e^{ - {\lambda _0}(t){T_{slot}}}} = 1 - } {e^{ - {\lambda _0}(t){T_{slot}}}} \\
 p_a^1(t) = {\lambda _1}(t){T_{slot}} \\
 \end{array} \right..
\label{eq32}
\end{equation}

Now, we have obtained 6 equations (consisting of Eqs. \eqref{eq23}, \eqref{eq29} and \eqref{eq31} ) including 8 unknown variables ${Ts_q}(t)$, $p_{s}^q(t)$, $\sigma _q(t)$ and $\rho _q(t)$ $(q=0,1)$. Next we use the following iterative method to calculate the unknown variables and further calculate the variance of service time ${Ds_{q}}(t)$ based on ${Ts_q}(t)$.

\begin{itemize}
\item[1)] Initialize $\rho_q^0(t)$;
\item[2)] Calculate the value of $p_{s}^q(t)$ and $\sigma _q(t)$ through solving Eqs. \eqref{eq29} and \eqref{eq31} based on the connectivity model $H(t)$, then calculate the mean of service time $Ts_q(t)$ by solving Eqs. \eqref{eq21}-\eqref{eq23} and Eqs. \eqref{eq25}-\eqref{eq27};
\item[3)] Calculate $\rho _q(t)$ according to the equation $min({\lambda_q}{Ts_q}(t), 1)$ and calculate the absolute error between the initial value of $\rho _q^0(t)$ and $\rho _{q}(t)$;
\item[4)] Compare the error with the predefined error bound $\varepsilon$. If the error is less than $\varepsilon$, the iteration is completed and the values of $p_{s}^q(t)$, $\sigma _q(t)$, ${Ts_q}(t)$ are obtained. Otherwise, set $\rho _{q}(t)$ as $\rho_q^0(t)$, then go to the second step to repeat the process of iteration.
\item[5)] Calculate the variance of service time ${Ds_{q}}(t)$ according to Eq. \eqref{eq24}
\end{itemize}

So far, we have obtained the mean and variance of service time at time $t$, i.e., $Ts_q(t)$ and ${Ds_{q}}(t)$, based on the connectivity model $H(t)$, thus the squared coefficient variation of service time $c_q(t)^2$ and average service rate $\mu_q(t)$ are determined accordingly. However, Eq. \eqref{eq18} and Eq. \eqref{eq19} are complicated and difficult to be analytically solved, thus we employ a numerical method to solve them. Specifically, given the initial average number of packets, i.e., $L_q(t_0)$, a constant arrival rate within the time interval $[t_0, t_0+\Delta t]$, i.e., $\lambda_q(t_0)$, and the calculated $c_q(t_0)$ and $\mu_q(t_0)$, we can use the Runge-Kutta algorithm to calculate the average number of packets at $t_0+\Delta t$, i.e., $L_q(t_0+\Delta t)$, and the change rate of $L_q(t_0+\Delta t)$, i.e., $\mathop {{L_q}}\limits^. (t_0+\Delta t)$. Afterward, we can repeat the above procedure to obtain ${L_q}$ and $\mathop {{L_q}}\limits^. $ in the following time interval, e.g., $[t_0+\Delta t, t_0+2\Delta t]$. Therefore, ${L_q}(t)$ and $\mathop {{L_q}}\limits^. (t)$ can be calculated by repeating the process described above.

\subsection{Performance Metrics}
In this section, we derive two important performance metrics of 802.11p, i.e., packet delay  and packet delivery ratio.
\subsubsection{Packet Delay}

Packet delay refers to the time interval from the time that a packet arrives at the transmission queue of the target vehicle to the time that the packet is received or dropped. Let $T_s^q$(t) be the average sojourn time of a packet in $AC_q$ at time $t$. According to the Little's law, $L_q(t)$ is equal to
\begin{equation}
L_q(t) = \lambda _q(t) \times T_s^q(t).
\label{eq33}
\end{equation}

Since ${\lambda _q}(t)$ is assumed to be a constant within each time interval $\Delta t$, thus the change rate of the average sojourn time can be calculated according to Eq. \eqref{eq33}, i.e.,
\begin{equation}
\mathop {{T_s^q}}\limits^.(t)  = \frac{{{{\mathop {L_q}\limits^. }}(t)}}{{{\lambda _q}(t)}}.
\label{eq34}
\end{equation}

The packet delay at time $t$ can be calculated as the summation of the packet delay at time $t- \Delta t$, i.e., ${PD_q}(t- \Delta t)$ and the change of the average sojourn time within time interval $[t - \Delta t,t]$ which is equal to the integral of $\mathop {T_s^q}\limits^.(t- \Delta t)$ within $[t - \Delta t,t]$. Thus the packet delay can be calculated as
\begin{equation}
P{D_q}(t) = P{D_q}(t - \Delta t) + \int_{t - \Delta t}^t {\frac{{\mathop {{L_q}}\limits^. (t-\Delta t )}}{{{\lambda _q}(t-\Delta t)}}dt,q = 0,1},
\label{eq35}
\end{equation}
where $\lambda _q(t-\Delta t)$ is given and $\mathop {L_q}\limits^.(t-\Delta t)$ is calculated through the numerically methods described at the end of Section \uppercase\expandafter{\romannumeral4-$C$}. The initial ${PD_q}(t_0)$ is calculated as
\begin{equation}
P{D_q}({t_0}) = \frac{{{L_q}({t_0})}}{{{\lambda _q}({t_0})}}.
\label{eq35}
\end{equation}

Given the calculated initial ${PD_q}(t_0)$, ${PD_q}(t- \Delta t)$ in Eq. \eqref{eq35} can be calculated through the numerical method. As a result, the time-varying packet delay of $AC_q$ can be calculated according to Eq. \eqref{eq35}.

\subsubsection{Packet Delivery Ratio}
Packet delivery ratio of $AC_q$ at time $t$ is defined as the ratio between the traffic successfully received by the vehicles within the transmission range of the target vehicle for the serving traffic of $AC_q$, which is denoted as $f_s^q(t)$, and the traffic that should be received by the vehicles for the arriving traffic of $AC_q$, which is denoted as $f_a^q(t)$, thus we have
\begin{equation}
PDR_q(t) = \frac{f_s^q(t)}{f_a^q(t)}.
\label{eq36}
\end{equation}

As the arrival rate at $AC_q$ is ${\lambda_q}(t)$ and $h_{i,j}^{k,l}$ indicates if a vehicle is within the transmission range of the target $V_{i,j}$, $f_a^q(t)$ can be expressed as
\begin{equation}
\begin{split}
f_a^q(t)&={\sum\limits_{{k} = 1}^n {\sum\limits_{l = 1}^m {\lambda_{q}(t){h_{i,j}^{k,l}(t)}}}}\\
&q = 0,1;\forall i,{k} = 1,2,...,n;\forall j,l = 1,2,..,m.
\end{split}
\label{eq37}
\end{equation}

Meanwhile, the served traffic of $AC_q$ at time $t$ is ${\mu_q}(t){\rho_q}(t)$, thus $f_s^q(t)$ is calculated as
\begin{equation}
\begin{split}
f_s^q(t)&={\sum\limits_{{k} = 1}^n {\sum\limits_{l = 1}^m {{\mu_q}(t){\rho_q}(t){h_{i,j}^{k,l}}}(t)\left[ {1 - {P_{{ij},{kl}}^c}(t)} \right]} } \\
&q = 0,1;\forall i,k  = 1,2,...,n;\forall j,{l}= 1,2,..,m;
\end{split}
\label{eq38}
\end{equation}
where ${P_{{ij},{kl}}^c}(t)$ is the collision probability when the target vehicle $V_{i,j}$ transmits to vehicle $V_{k,l}$. The collision occurs under two cases. One case is that at least one AC queue of another vehicle in the transmission range of $V_{i,j}$, i.e., exposed terminal, is transmitting at the same time. Let $\sigma _q^{u,w}(t)$ be the transmission probability of $AC_q$ of vehicle $V_{u,w}$, thus the collision probability caused by the exposed terminal is expressed as
\newcounter{TempEqCnt2}
\setcounter{TempEqCnt2}{\value{equation}}
\setcounter{equation}{40}
\setcounter{equation}{\value{TempEqCnt2}}
\begin{equation}
\begin{split}
&P_{ij,kl}^{exposed}(t) = {h_{i,j}^{k,l}} \\
&\times \left[ {1 - \prod\limits_{u = 1}^n {\prod\limits_{w = 1}^m {{{\left[ {(1 - \sigma _0^{u,w}(t))(1 - \sigma _1^{u,w}(t))} \right]}^{{h_{i,j}^{u,w}}(t)}}} } } \right],\\
&\forall i,k,u= 1,...,n;\forall j,l,w  = 1,...,m.
\end{split}
\label{eq39}
\end{equation}
\begin{equation}
\begin{split}
P_{{ij},{kl}}^{hidden}(t) = {h_{i,j}^{k,l}}(t) &\times \left[ {1\left. { - \prod\limits_{u = 1}^n {\prod\limits_{w = 1}^m {{{\left[ {(1 - \sigma _0^{u,w}(t))(1 - \sigma _1^{u,w}(t))} \right]}^{\frac{{{2T_{tr}}}}{{{T_{slot}}}}(1 - {h_{i,j}^{u,w}}(t)){h_{k,l}^{u,w}}(t)}}} } } \right]} \right.\\
&\forall  i,k,u  = 1,2,3,..,n;\forall j,l,w  = 1,2,3,...,m.
\end{split}
\label{eq40}
\end{equation}

Another case is caused by the hidden terminal, i.e., the vehicle which is in the transmission range of $V_{k,l}$ but not in the transmission range of $V_{i,j}$. Since $V_{i,j}$ is transmitting data to $V_{k,l}$ at time $t$ and the transmission delay is $T_{tr}$, a collision occurs at $V_{k,l}$ when at least one AC queue of a hidden terminal is transmitting to $V_{k,l}$ within the time interval $[t-T_{tr},t+T_{tr}]$. To avoid the collision, each AC of the hidden terminal should not transmit within each time slot of the time duration $2T_{tr}$ with probability ${\left[ {(1 - \sigma _0^{u,w})(1 - \sigma _1^{u,w})} \right]^{\frac{{{2T_{tr}}}}{{{T_{slot}}}}}}$. Hence, the collision probability caused by the hidden terminal is given by Eq. \eqref{eq40}, which is shown at the top of next page.


Since the collision may be incurred by the above two cases, ${P_{{ij},{kj}}^c}(t)$ can be derived as
\setcounter{equation}{41}
\begin{equation}
P_{ij,kl}^c(t) \!=\! 1\! -\! ( {1\! -\! P_{ij,kl}^{exposed}(t)})( {1\! -\! P_{ij,kl}^{hidden}(t)} ).
\label{eq41}
\end{equation}

Substituting Eqs. \eqref{eq39} and \eqref{eq40} into Eq. \eqref{eq41}$, {P_{{ij},{kj}}^c}(t)$ in Eq. \eqref{eq38} can be determined according to $\sigma _0(t)$, $\sigma _1(t)$ and $H(t)$.
As a result, packet delivery ratio of $AC_q$ at time $t$ can be determined according to Eq. \eqref{eq36}.

{It is worthy noting that compared to the highway scenarios, more factors should be considered when there is a bend. The whole bending process can be divided into three parts: before the bend, during the bend and after the bend. The calculation of the position of the vehicle is different in the different durations. The change of the position of vehicles would further affect the performance of 802.11p. Therefore, the analysis for the time-dependent performance of 802.11p when there is a bend is more complicated than the highway case. But the developed analysis method can be employed with similar way.}

\section{{Calculation Steps}}
In this section, we describe the algorithm to calculate the {time-dependent} 802.11p performance at any time $t$ $(t \in [{t_0},{t_f}])$ according to our model. The detailed procedure of the algorithm is shown as follows.

\emph{Step 1}: Initialize the kinestate of each vehicle, average number of packets and packet arrival rate, i.e., $(x(t_0),y(t_0))$, $v(t_0)$, $a(t_0)$, ${L_q}({t_0})$ and ${\lambda_q}({t_0})$. Then, calculate $H(t_0)$ based on the kinestate of each vehicle through Eqs. \eqref{eq3}-\eqref{eq4} and ${PD_q}({t_0})$ based on ${L_q}({t_0})$ and ${\lambda_q}({t_0})$ through Eq. \eqref{eq35'} .

\emph{Step 2}: Calculate the transmission probability and the mean and variance of service time of each vehicle at $t_0$, i.e., ${\sigma_q}(t_0)$, ${Ts_q}(t_0)$ and ${Ds_q}(t_0)$, by solving Eqs. \eqref{eq21}-\eqref{eq32} with iterative method described in Section \uppercase\expandafter{\romannumeral4-$C$}. Calculate the squared coefficient variation of service time and average service rate at time $t_0$, i.e., $c_q(t_0)^2$ and $\mu_q(t_0)$, based on the obtained ${Ts_q}(t_0)$ and ${Ds_q}(t_0)$ through Eqs. \eqref{eq20} and \eqref{eq20'}. Set $t=t_0$.

\emph{Step 3}: Calculate ${L_q}({t} + \Delta t)$ and $\mathop {{L_q}}\limits^. ({t} + \Delta t)$ based on ${L_q}({t})$, $c_q(t)^2$ and $\mu_q(t)$ by solving Eqs. \eqref{eq18} and \eqref{eq19} with the Runge-Kutta algorithm. Then, the calculated ${L_q}({t} + \Delta t)$ becomes the initial value of next time interval $[{t} + \Delta t,{t} + 2\Delta t]$. Afterwards, calculate ${\rho_q}({t} + \Delta t)$ based on ${L_q}({t} + \Delta t)$ through Eqs. \eqref{eq15} and \eqref{eq17}.



\emph{Step 4}: Calculate kinestate of each vehicle at $t+\Delta t$, i.e., $x(t+\Delta t)$,  $v(t+\Delta t)$ and $a(t+\Delta t)$, based on $x(t)$, $v(t)$ and $a(t)$ through Eqs. \eqref{eq5}-\eqref{eq12}, and then calculate $H(t+\Delta t)$ based on $x(t+\Delta t)$  through Eqs. \eqref{eq3}-\eqref{eq4}.

\emph{Step 5}: Given ${\lambda_q}(t+\Delta t)$, calculate ${\sigma_q}(t+\Delta t)$, ${Ts_q}(t+\Delta t)$ and ${Ds_q}(t+\Delta t)$ based on $H(t+\Delta t)$ through Eqs. \eqref{eq21}-\eqref{eq32} and then calculate $c_q(t+\Delta t)^2$ and $\mu_q(t+\Delta t)$ based on the ${Ts_q}(t+\Delta t)$ and ${Ds_q}(t+\Delta t)$ through Eqs. \eqref{eq20} and \eqref{eq20'}.

\emph{Step 6}:  Calculate packet delay at time ${t} + \Delta t$, i.e., ${PD_q}({t} + \Delta t)$, based on ${\lambda_q}(t)$, $\mathop {{L_q}}\limits^. (t)$ and ${PD_q}({t})$ through Eq. \eqref{eq35}.

\emph{Step 7}: Calculate the collision probability ${P_{ij,kl}^{c}}({t} + \Delta t)$ based on the obtained $H({t} + \Delta t)$ and ${\sigma_q}({t} + \Delta t)$ through Eqs. \eqref{eq39}-\eqref{eq41}, then calculate the packet delivery ratio, i.e., ${PDR_q}({t} + \Delta t)$, based on ${P_{ij,kl}^{c}}({t} + \Delta t)$, ${\rho_q}({t} + \Delta t)$ and $H({t} + \Delta t)$ through Eqs. \eqref{eq36}-\eqref{eq38}.

\emph{Step 8}: Set $t=t+\Delta t$. If $t< t_f$, change the time interval $[t,t+\Delta t]$ to the next time interval $[t+\Delta t,t+2\Delta t]$ and return to Step $3$. Otherwise, terminate the iteration.

\section{Model Validation and Performance Analysis}

\begin{table}
\scriptsize
\caption{Related Parameter Values}
\label{tab2}
\centering
\begin{tabular}{cccc}
\toprule
\textbf{Parameter} &\textbf{Value} &\textbf{Parameter} &\textbf{Value}\\
\midrule
$a$ & $1.4m/{s^2}$ & $b$ & $2m/{s^2}$\\
\midrule
${s_0}$ & $3m$ &${v_0}$ & $30m/s$\\
\midrule
${T_m}$ & $1.5s$ &$ {T_l} $ &$2s$\\
\midrule
${t_d}$ & $10s$ & $ {t_s} $ &$10s$\\
\midrule
$R$ & $2$ & $ {R_{tr}} $ &500m\\
\midrule
$ m$ & $8$ & $n$ & $9$\\
\midrule
$v_{stb}$ & $25m/s$ & $v_{low}$ & $5m/s$\\
\midrule
$\delta$  & $2 \mu s$ & ${T_{slot}}$ & $13 \mu s$\\
\midrule
$\Delta t$ & $0.01s$ & $CW_{0,min}$ &$3$\\
\midrule
$CW_{1,min}$ & $3$ & $CW_{1,max}$ &$7$\\
\midrule
$AIFSN_0$ & $2$ & $AIFSN_1$ &$3$\\
\midrule
$R_d$ & $6Mbps$ & $R_b$ &$1Mbps$\\
\midrule
$PH{Y_H}$ & $48bits$ & $MA{C_H}$ &$112bits$\\
\midrule
$SIFS$ & $32\mu s$ & $E[P]$ & $200bits$\\
\bottomrule
\end{tabular}
\end{table}

In this section, we validate the effectiveness of the constructed models through simulation results. The simulation is conducted on MATLAB 2010a.
The simulation scenario is described in the system model and the initial scenario is shown in Fig.\ref{fig1}.
We consider that the target vehicle is $V_{2,1}$ which is also the disturbed vehicle and each vehicle has two ACs to broadcast packets. {and use the velocity change of a vehicle under disturbance proposed in [8] and [30] to describe the time-varying kinestate of $V_{i^*,j^*}$, which is shown in Fig. 2. {The disturbance may be caused by the leader vehicle acceleration/deceleration, wind gust and uncertainties in a platoon control system consisting of the aerodynamics drag, rolling resistance moment, random non-Gaussian and non-linear noise, and variation of vehicle mass} \cite{Guo2011, Liu2014}.}
The initial time is $0s$.
The width of each lane is $3.5m$ and the vehicle length is $3m$. The parameters of 802.11p are configured according to the IEEE 802.11p standard \cite{Hafeez2015}. The values of the related parameters are shown in Table \ref{tab2}, where $T_l$ and $T_m$ are the desired headway for the lead vehicle and following vehicles, respectively. Moreover, we adopt the analytical results to analyze the real-time performance of 802.11p for a target vehicle after a vehicle in platoons suffers from the disturbance.
\begin{figure}
\centering
\includegraphics[scale=0.8]{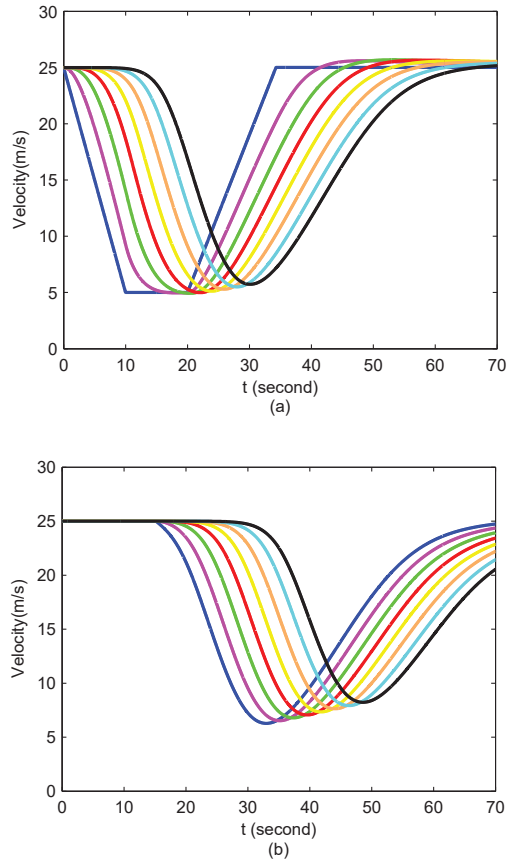}
\caption{{Time-dependent} velocity of the vehicles in platoons $P_2$ and $P_3$. (a) The curves represent the {time-dependent} velocity of $V_{2,1},\ldots,V_{2,8}$ from left to right, respectively. (b) The curves represent the {time-dependent} velocity of $V_{3,1},\ldots,V_{3,8}$ from left to right, respectively.}
\label{fig10}
\end{figure}
Fig. \ref{fig10} shows the {time-dependent} velocity of the vehicles in platoons $P_2$ and $P_3$ where majority of the vehicles are in the transmission range of the target vehicle $V_{2,1}$. It can be seen that vehicles in $P_2$ and $P_3$ change their velocities one by one according to the IDM model after the target vehicle $V_{2,1}$ changes its velocity at the initial time $0$.

\begin{figure}[!htbp]
\centering
\includegraphics[scale=0.74]{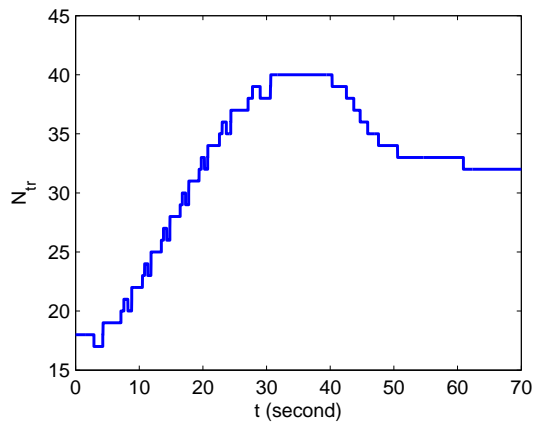}
\caption{{Time-dependent} number of vehicles in the transmission range of target vehicle $V_{i,j}$.}
\label{fig11}
\end{figure}

Fig. \ref{fig11} shows the {time-dependent} number of vehicles within the transmission range of the target vehicle, i.e., $N_{tr}(t)$. It can be seen that $N_{tr}(t)$ steps up basically before $30.6s$. This is because that majority of the following vehicles in $P_2$ and $P_3$ decelerate their velocities during $[0,30.6]s$, which reduces the corresponding inter-platoon spacing and intra-platoon spacing and thus increases $N_{tr}(t)$. However, it can be seen that $N_{tr}(t)$ sometimes steps down before $30.6s$. This is because that the vehicles in front of the target vehicle on a common lane keep moving with $v_{stb}$ while the velocity of the target vehicle is smaller than $v_{stb}$ before $30.6s$. In this case, the distance between the front vehicles and the target vehicle increases and thus the number of the front vehicles in the transmission range of the target vehicle is decreased. As a result, the total number of vehicles in the transmission range of the target vehicle is sometimes decreased. Moreover, it can be seen that $N_{tr}(t)$ keeps constant in $[30.6,40.3]s$. This is because that a part of the following vehicles accelerate and another part of the following vehicles decelerate as shown in Fig. \ref{fig10}. Therefore, the summation of the corresponding inter-platoon spacing and intra-platoon spacing almost keeps a constant and leads to a constant $N_{tr}(t)$ within $[30.6,40.3]s$. In addition, it can be seen that $N_{tr}(t)$ steps down after $40.3s$. This is because majority of the following vehicles accelerate after $40.3s$, thus increasing the corresponding intra-platoon spacing and inter-platoon spacing.

\begin{figure}
\centering
\includegraphics[scale=0.8]{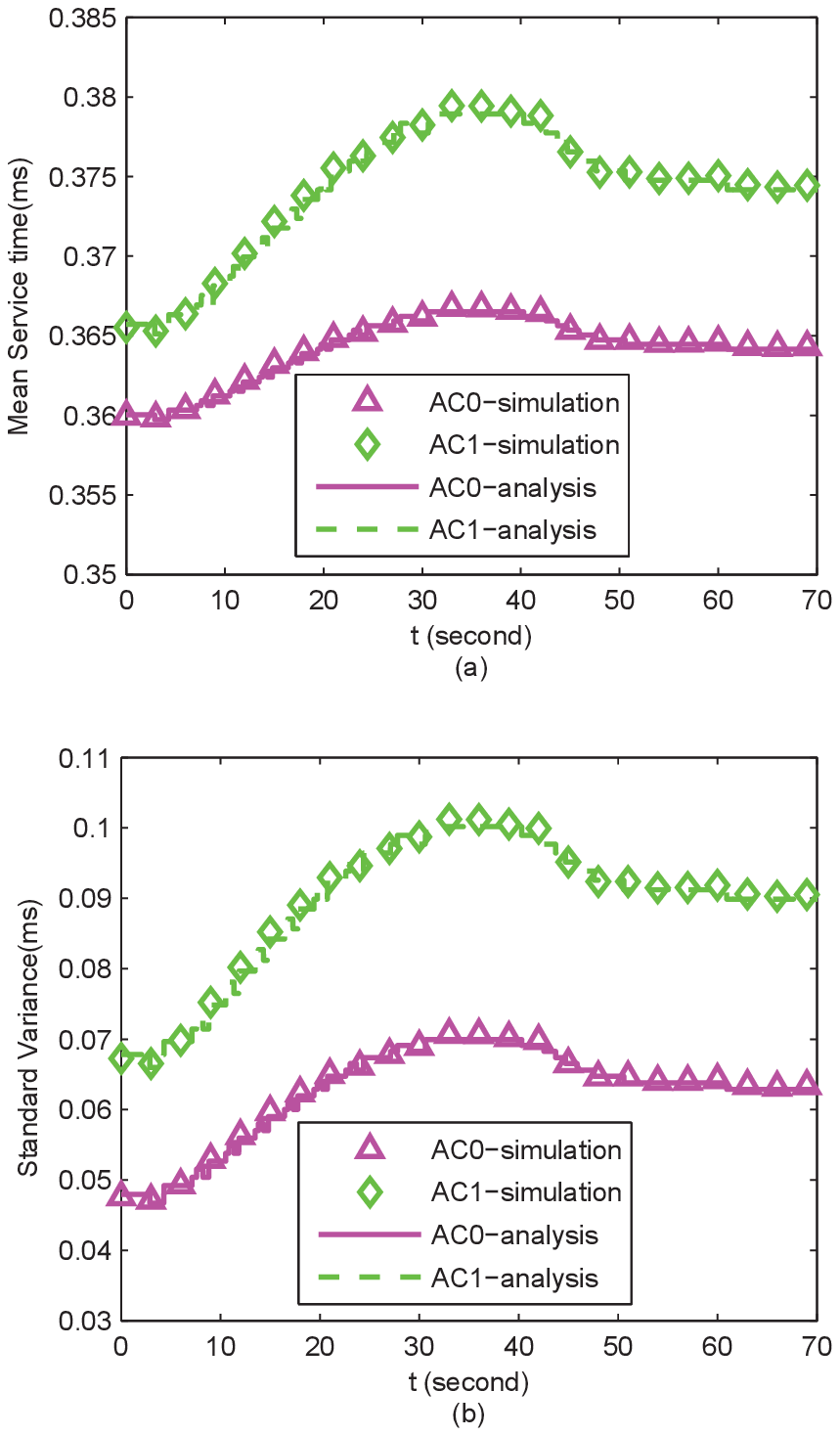}
\label{fig:side:a}
\caption{{Time-dependent} mean and standard variance of service delay ($\lambda_0=\lambda_1=20 pkt/s$). (a) {Time-dependent} mean service delay. (b) {Time-dependent} standard of service time.}
\label{fig12}
\end{figure}

{Fig. \ref{fig12} shows} the {time-dependent} mean and standard variance of service time for each AC of the target vehicle, i.e., $Ts_q(t)$ calculated by Eq. \eqref{eq23} and the square root of $Ds_q(t)$ calculated by Eq. \eqref{eq24}. It is seen that the simulation results are very close to the analysis results. Moreover, the trends of both the {time-dependent} mean and standard variance of the service time are similar to that of the time-varying number of vehicles within the transmission range. This is because that $Ts_q(t)$ and $Ds_q(t)$ are calculated based on $N_{tr}(t)$ according to Eqs. \eqref{eq21}-\eqref{eq32}. Moreover, the mean and standard variance of service delay for $AC_0$ $(q=0, 1)$ are lower than $AC_1$. This is because that $AC_0$ is granted higher priority than $AC_1$ and has the lower mean and standard variance of the service delay.

\begin{figure*}
\centering
\includegraphics[scale=0.73]{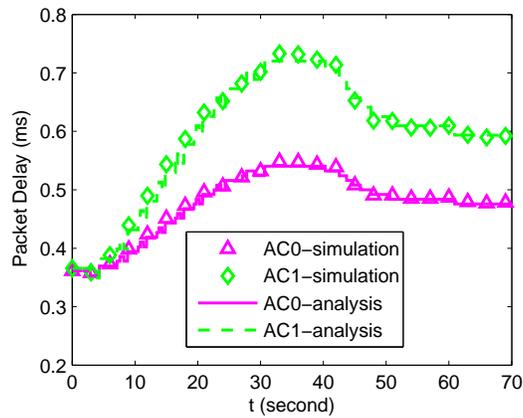}
\caption{{Time-dependent} packet delay ($\lambda_0=\lambda_1=20 pkt/s$).}
\label{fig14}
\end{figure*}

Fig. \ref{fig14} shows the {time-dependent} packet delay for each AC of the target vehicle, i.e., $PD_q(t)$ calculated by Eq. \eqref{eq35}. It is seen that the simulation results are very close to the analytical results. Moreover, the trend of {time-dependent} packet delay is consistent with that of the number of vehicles within the transmission range of the target vehicle. This is because that $PD_q(t)$ is calculated based on $N_{tr}(t)$ according to Eqs. \eqref{eq18}-\eqref{eq35'}. Moreover, it can be seen that both of the maximum {time-dependent} packet delays of $AC_0$ and $AC_1$ are smaller than the time interval $\Delta t$, i.e., $0.01s$, which indicates that both the event-driven information and periodic information can be received by other vehicles in time to make the reaction in advance.

\begin{figure}
\centering
\includegraphics[scale=0.72]{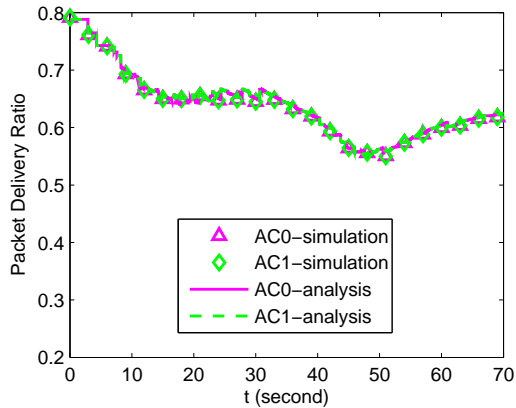}
\caption{{Time-dependent} packet delivery ratio ($\lambda_0=\lambda_1=20 pkt/s$).}
\label{fig15}
\end{figure}

Fig. \ref{fig15} shows the {time-dependent} packet delivery ratio for each AC of the target vehicle, i.e., $PDR_q(t)$ calculated by Eq. \eqref{eq36}. It is seen that the simulation results are very close to the analytical results. Moreover, the packet delivery ratio is large initially, then it decreases dramatically before $14.3s$ and increases slightly during the period $[14.3;30.6]s$, decreases with little jitter within $[30.6;40.3]s$ and $[40.3;50.6]s$ with dramatical jitter. Finally increases slightly after $50.6s$. The reason is explained as follows. Initially, $N_{tr}(t)$ is small and few vehicles contend to transmit data.
Thus the collision probability is small and incurs a high packet delivery ratio. Then more following vehicles enter into the transmission range of the target vehicle and thus incur the increasing number of hidden-terminals, which results in the packet delivery ratio decreases dramatically before $14.3s$. After that, the hidden terminals move into the transmission range and become the exposed terminals, thus leading to the decreasing number of the hidden terminals. Therefore, the packet delivery ratio is improved during the period $[14.3,30.6]s$. Afterwards, according to Fig. \ref{fig11}, $N_{tr}$ reaches the maximum value and does not change within $[30.6,40.3]s$, but the velocity of most of the following vehicles within the transmission range of the target vehicle are smaller than that of the vehicles moving on the other lanes, i.e., $v_{stb}$. So more vehicles on other lanes move into the transmission range of the following vehicles and become the hidden terminals of the target vehicle, leading to the packet delivery ratio is decreased. Moreover, according to Fig. \ref{fig10}, the velocities of the following vehicles do not change dramatically within $[30.6,40.3]s$, thus the packet delivery ratio is decreased with little jitter. Afterwards, during $[40.3,50.6]s$, most of the following vehicles are accelerating and the inter-platoon spacing and intra-platoon spacing are increasing accordingly, resulting in that these vehicles move out of the transmission range of the target vehicle and become the hidden terminals, and further decrease the packet delivery ratio with a dramatical jitter. Finally, after $50.6s$, the velocities of all the following vehicles are further increased and the corresponding intra-platoon spacings and inter-platoon spacings are further increased, which reduces the hidden-terminals. Therefore the packet delivery ratio is improved after $50.6s$. Moreover, the time-varying packet delivery ratio of $AC_0$ and $AC_1$ is almost the same due to the similar collision probability for two ACs. {We can see that the time-dependent PDR is relatively high, which means most of packets can be received successfully, thus the communication requirement can be guaranteed.}

Finally, we adopt the maximum deviation between the simulation results and the analytical results to describe the accuracy of the constructed models. Specifically, Maximum Deviation = Max (($|$ Simulation results - Analytical results $|$) $/$ Analytical results $\times$ $ 100 \% $). The maximum deviations under different conditions are shown in Table \ref{tab3}. It is seen that the accuracy of the constructed models is high and the maximum deviations under different conditions are less than 3\%.

\begin{table}
\scriptsize
\caption{Accuracy of the Model}
\label{tab3}
\centering
\begin{tabular}{ccc}
\toprule
\textbf{Performance Metric} &\textbf{ACs} &\textbf{Max. Deviation}\\
\midrule
$PD$ & $0$ & $1.72\% $\\
\midrule
$PDR$ & $0$ &$1.54\%$\\
\midrule
$PD$ & $1$ & $2.80\% $\\
\midrule
$PDR$ & $1$ &$1.62\%$\\
\bottomrule
\end{tabular}
\end{table}

\section{Conclusion}
In this paper, we have constructed models to analyze the {time-dependent} performance of 802.11p for platooning communications under the disturbance. A time-varying connectivity model was first constructed based on the kinestate of each vehicle under the disturbance to reflect the dynamic connectivity among vehicles. Then, the PSFFA approach was adopted to model the dynamic queuing behavior of different ACs and derive the {time-dependent} average number of packets in each AC. Afterwards, the mean and variance of the service time of each AC were derived based on the connectivity model according to the access process of 802.11p to obtain the time-varying average number of packets in each AC. Finally, the {time-dependent} performance of each AC in terms of the packet delay and packet delivery ratio has been derived for the target vehicle. The effectiveness of the models were validated through simulation results. Moreover, the numerical results were used to analyze the dynamic performance of the 802.11p and validated that 802.11p is able to maintain the string stability under disturbance. {For the future work, we will take the physical channel model into account to analyze the time-dependent performance of 802.11p-based platooning communications and also discuss the cases where vehicles in the platoon need to change lane or encounter a bend.}



\ifCLASSOPTIONcaptionsoff
  \newpage
\fi

%
%
%
\bibliographystyle{unsrt}
\bibliography{IEEE_TVT}

\end{spacing}
\end{document}